# Optimal Decision Making Model of Battery Energy Storage-Assisted Electric Vehicle Charging Station Considering Incentive Demand Response

Bishal Upadhaya[1], Donghan Feng[1], Yun Zhou[1*], Qiang Gui[1], Xiaojin Zhao[1], Dan Wu[2]

[1] Key Laboratory of Control of Power Transmission and Conversion, Ministry of Education, Department of Electrical Engineering, Shanghai Jiao Tong University, 800 Dongchuan Rd., Shanghai, China
[2] Marketing Department, State Grid Shanghai Municipal Electric Power Company, 1122 Yuansheng Rd., Shanghai, China
*yun.zhou@sjtu.edu.cn

**Keywords**: Optimal decision making model, Electric vehicle charging station, Battery energy storage, Emergency demand response

## Abstract

Considering large scale implementation of electric vehicles (EVs), public EV charging stations are served as fuel tanks for EVs to meet the need of longer travelling distance and overcome the shortage of private charging piles. The allocation of local battery energy storage (BES) can enhance the flexibility of the EV charging station. This paper proposes an optimal decision making model of the BES-assisted EV charging station considering the incentive demand response. Firstly, the detailed models of the BES-assisted EV charging station are presented. Secondly, as a representative incentive demand response, the emergency demand response (EDR) model is introduced. Thirdly, based on the charging load forecast data, an optimal decision making model of the BES-assisted EV charging station considering the EDR to maximize the charging station's operating profit is established. Finally, the feasibility of the proposed method is verified through case studies. The conclusions of this paper are as follows: 1) Through the optimal decision making model, correct and profitable EDR participation decision can be determined for the BES-assisted EV charging station effectively. 2) Local BES in the EV charging station can improve the charging station's ability to participate in the EDR.

## 1 Introduction

Governmental policies and incentives have been beneficial for the adoption of electric vehicles (EVs) all around the world [1]. In a 2018 report [2], International Energy Agency (IEA) predicted an astronomical surge in EVs on the road to 130 million by 2030 and may even be higher to 228 million under the best-case scenario. Even though higher penetration of EVs bring down the greenhouse gases emission but also increases noticeable power demand on the power system [3]. The increased load must be dealt economically either by adding power generation or effective demand-side management techniques (DSMT) for sound operation of power system. With the advancement in smart grid technologies, demand response (DR) is one of the most promising DSMT to manage the electrical load economically [4]. Studies [5]-[6] illustrate the potential of EVs participation in demand response.

The U.S Department of Energy (DOE) describes the demand response as: Changes in electric usage by end-use customers from their normal consumption patterns in response to changes in the price of electricity over time, or to incentive payments designed to induce lower electricity use at times of high wholesale market prices or when system reliability is jeopardized [7]. DOE classifies DR programs into two categories namely, price-based DR programs and incentive-based DR programs. One of the incentive-based DR programs is emergency demand response (EDR) which is called upon by independent system operator (ISO) during the emergency event. EDR participation is voluntary if consumers opt to participate in energy only EDR scheme while in full EDR scheme, and it is mandatory to shed the committed capacity during EDR event otherwise penalty will be charged for non-participation [8]-[10]. In PJM, EDR has gained popularity because of its incentive scheme both for capacity and performance during DR event, which can be seen by its load reduction commitment capacity inflation over time from 1700MW in year 2006/2007 to 14800MW in year 2015/2016 [11]. With the EVs expected to grow in the coming days [2], the EV charging station not only play the role of an aggregator of EVs but also provides a platform for distributed resources to integrate, which makes EV charging station an important EDR resource.

Many research papers have pointed out the benefits of EDR in spectra of power system application for reducing system cost and peak demand to establish reliability during emergency event. In papers [12]-[13], technical and economic advantages of EDR program on electrical load in Iranian power grid is presented. In [14], assessment of EDR in distribution system to pacify the overloading of specific overloaded node is studied. In [15], a coordinated energy management technique in mixed use building is proposed to enable EDR program. With EDR program, the system cost is reduced significantly. In [16], optimal scheduling of chillers and energy storage system in a commercial building considering EDR is presented.

While the majority of research papers have focused on EV charging station participation in price-based DR (especially TOU) to alleviate grid peak along with minimizing EV charging station cost [17]- [19], only very few research works have been carried out that focuses on both price based and incentive-based DR participation of EV charging station. In paper [20], both price-based and incentive-based DRs are considered and their effects on the operational behaviour of EV parking load under uncertainties is studied. In [21], a real time EV charging scheduling scheme is proposed to minimize the EV charging cost considering dynamic tariff and load reduction DR signal.

In paper [22]-[23], the flexibility and economics of incorporating BES in charging station to reduce stress on the power grid during peak load hours under TOU price system is studied. In a report [24] by BloombergNEF, cost of battery storage system (BES) is predicted to fall as low as $70/kWh by 2030 which could be a driving force for adopting BES to add flexibility to EV charging station. With BES allocated in the EV charging station, the BES-assisted EV charging station becomes more flexible. And the BES-assisted EV charging station can participate in the EDR effectively.

However, to participate in the EDR, the EV charging station should drop its net load supplied by the grid down to the EDR baseline provided by the ISO. The EV charging load curtailment will happen due to the decrease of the power supplied by the grid. The main driving force for the BES-assisted EV charging station to participate in the EDR is that the potential profit loss because of EV charging load curtailment may be under-compensated, compensated, or over-compensated by the incentive payment for EDR participation from the ISO. For the over-compensated scenario, it's profitable for the BES-assisted EV charging station to participate in the EDR.

In this paper, an optimal decision making model of the BES-assisted EV charging station considering the EDR to maximal the charging station's operating profit is proposed. Detailed optimal scheduling model for the BES-assisted EV charging station considering EDR and the EDR participation decision making process for the BES-assisted EV charging station considering EDR is established in this paper. Through the EDR participation decision making process, the decision for the EV charging station whether to participate in the EDR can be determined effectively. And for the EDR participation case, the optimal scheduling for the BES-assisted EV charging station during the EDR event duration can be determined meanwhile.

The rest of the paper is organized as follows. The methodology for BES-assisted EV charging station participated in EDR is explained in Section 2. Section 3 presents the detailed optimal decision making model for BES-assisted EV charging station participated in EDR. Case studies to demonstrate the proposed model are included in Section 4. Eventually, Section 5 concludes the paper.

## 2 Methodology for BES-assisted EV charging station participated in EDR

As flexible customers in the power system, the BES-assisted EV charging station can participate in the EDR event potentially. The methodology for BES-assisted EV charging station participated in EDR is presented in this section, including the time interval chart of the BES-assisted EV charging station participated in EDR and the information interaction between ISO and BES-assisted EV charging station during EDR

### 2.1 Time interval chart of the BES-assisted EV charging station participated in EDR

In this paper, the BES-assisted charging station is considered to participate in the EDR. The charging station purchases electrical power in the wholesale electricity market to satisfy charging station demand (i.e., EVs and BES charging demand). In the proposed model, the charging station is assumed to have signed up for energy only EDR scheme and hence, charging station response to DR signal is voluntary. During the EDR event, grid operator dispatches the load reduction signal along with payment information which must be responded within two hours of notification by the DR resource [8]-[10]. The short lead time resources request response decision time less than 1 hour. And for long lead time resource, the response decision time is requested within (1 hour, 2 hour]. As a typical short lead time resource, after notification of EDR, the decision making duration for the BES-assisted EV charging station whether to participate in the EDR is usually set less than 1 hour.

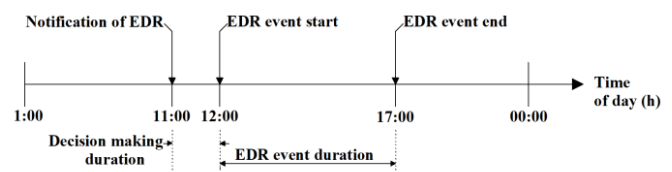

Fig. 1 Schematic time interval chart of the EV charging station responds to an EDR event.

The schematic time interval chart of the EV charging station responds to an EDR event is shown in Fig. 1. The notification of EDR for the EV charging station is set to 11:00 for example in Fig. 1. Considering the decision making duration to decide whether to participate in the EDR (i.e. 1 hour), the EDR event starts from 12:00. An event of EDR usually lasts for 2-6 hours or even longer [8], [10]. In this paper, the EDR lasting for 5 hours is considered for the EV charging station. Shown in Fig. 1, the EDR event duration starts at 12:00 and ends at 17:00.

*2.2 Information interaction between ISO and BES-assisted EV charging station during EDR*

Fig. 2 shows the information interaction between ISO and BESS-assisted EV charging station during EDR. In Fig. 2, the EV charging station short-term forecast load is provided by the BES-assisted EV charging station for the ISO periodically. If the ISO decides to offer an EDR event (e.g. at 11:00 in Fig. 2), it will dispatch different load reduction signals for different customers with the incentive payment for EDR participation according to the entire power system's state and the customer's short-term forecast load. The EDR baseline signal provided by the ISO is determined through the difference of the short-term forecast load and the load reduction signal.

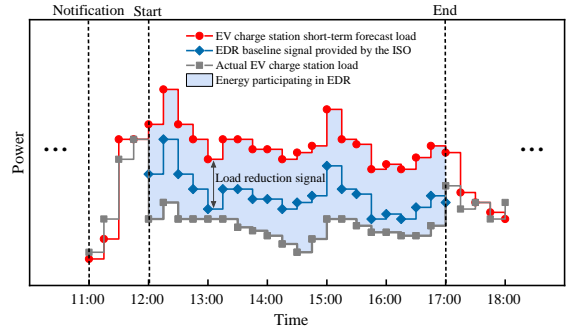

Fig. 2 Information interaction between ISO and BESS-assisted EV charging station during EDR

In the decision making duration (i.e. 11:00-12:00 in Fig.1 and 2), decision should be made by the BES-assisted EV charging station on whether to participate in the EDR. If the BES-assisted EV charging station confirms to participate in the EDR, the actual EV charging station net load supplied by the grid should not exceed the EDR baseline signal during the EDR event duration (i.e. 12:00-17:00 in Fig.1 and 2). And the shaded area in Fig. 2 (integration of the difference between the short-term forecast load and the actual load of the EV charging station) is the total energy of the BES-assisted EV charging station participated in the EDR and is the basis for accounting the total EDR incentive payment by the ISO to the BES-assisted EV charging station.

## 3 Optimal decision making model for BES-assisted EV charging station participated in EDR

The methodology for BES-assisted EV charging station participated in EDR is presented in Section 2. As shown in Fig. 2, to participate in the EDR, the EV charging station should drop its net load supplied by the grid down to the EDR baseline provided by the ISO. The EV charging load curtailment will happen due to the decrease of the power supplied by the grid. The mainly driving force for the BES-assisted EV charging station to participate in the EDR is that the potential profit loss because of EV charging load curtailment may be under-compensated, compensated, or over-compensated by the incentive payment for EDR participation from the ISO. For the over-compensated scenario, it's profitable for the BES-assisted EV charging station to participate in the EDR.

Referring to Fig. 1 and 2, in the decision making duration, the BES-assisted EV charging station should make the decisions on whether to participate in the EDR or not. The optimal scheduling model for the BES-assisted EV charging station considering EDR is established in this section, and the EDR participation decision making process for the BES-assisted EV charging station considering EDR is also presented in this section.

*3.1 Formulations of the optimal scheduling model for the BES-assisted EV charging station considering EDR*

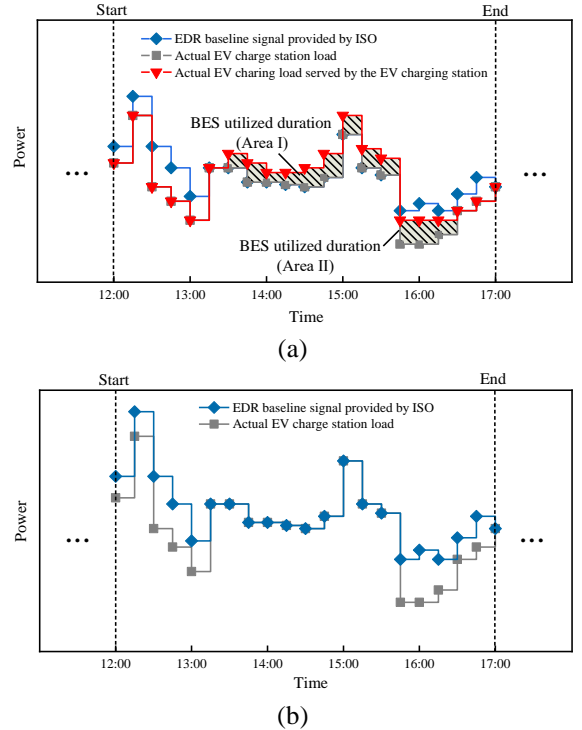

Fig. 3 Schematic diagram of the BES utilized durations during the EDR event duration

Considering the load reduction signal shown in Fig.2, if the BES-assisted EV charging station confirms to participate in the EDR, a minimum decrease in the net load supplied by the grid (i.e. the load reduction signal from the ISO) is requested to obtain the active load during the EDR event duration under the EDR baseline signal. As illustrated in Fig. 3, with the help of the BES allocated in the EV charging station, the actual EV charging power served by the EV charging station can exceed the EDR baseline limit, while the threshold-crossing part must be compensated by the BES to maintain the actual load of the EV charge station under the EDR baseline limit (e.g. Area I in Fig. 3). Except for the compensation of the threshold-crossing part, the abundant energy stored in the BES can be used to supply the EV charging load during the EDR event duration (e.g. Area II in Fig. 3).

The EDR participation decision making optimization model for the BES-assisted EV charging station is formulated with objective function and constraints as follows.

A. Objective function

The objective of the optimal scheduling model for the BES-assisted EV charging station considering EDR is to maximize the operating profit of the EV charging station $C_{EDR}$. The objective function of the optimization model is shown in (1).

$$\max C_{EDR} = \sum_{t \in S_H} \rho_{EV}(t) P_{EV}(t) \Delta t + \sum_{t \in S_H} \rho_{EDR}(t) \Delta P_{EDR}(t) \Delta t \\ - \sum_{t \in S_H} \rho_{Grid}(t) P_{Load}^{EDR}(t) \Delta t \quad (1)$$

$$S_H = \{t_s, t_s+1, ..., t_s + H/\Delta t - 1\} \quad (2)$$

The first item in the objective function $\rho_{EV}(t) P_{EV}(t) \Delta t$ is the income from the EV charging loads, where $P_{EV}(t)$ is the actual EV charging load at discrete time step $t$, $\rho_{EV}(t)$ is the time of use (TOU) charging price for the EVs, and $\Delta t$ is the time step width. The second item $\rho_{EDR}(t) \Delta P_{EDR}(t) \Delta t$ is the incentive payment for EDR participation from the ISO, where $\Delta P_{EDR}(t)$ is the decreased load amount that participated in the EDR, and $\rho_{EDR}(t)$ is the EDR participation incentive payment price provided by the ISO. The third item $\rho_{Grid}(t) P_{Load}^{EDR}(t) \Delta t$ is the cost for the electricity purchased from the grid, where $P_{Load}^{EDR}(t)$ is the actual EV charging station net load supplied by the grid during the EDR event duration, and $\rho_{Grid}(t)$ is the TOU electricity price of the grid. The set of discrete time steps in the EDR event duration $S_H$ is defined in (2), where $t_s$ is the starting time step in the EDR event duration, and $H$ is the time horizon of the EDR event duration.

B. Constraints

The constraints considered in the optimization model are shown in (3)-(14) below.

$$\Delta P_{EDR}(t) = P_{Load}^{Fore}(t) - P_{Load}^{EDR}(t) \quad t \in S_H \quad (3)$$

$$\Delta P_{EDR}(t) \geq \Delta P_{EDR}^{\min}(t) \quad t \in S_H \quad (4)$$

$$P_{Load}^{EDR}(t) \geq 0 \quad t \in S_H \quad (5)$$

$$P_{EV}(t) = P_{Load}^{EDR}(t) + P_{BES}(t) \quad t \in S_H \quad (6)$$

$$P_{BES}(t) = P_{BES}^{dis}(t) - P_{BES}^{ch}(t) \quad t \in S_H \quad (7)$$

$$0 \leq P_{BES}^{dis}(t) \leq P_{BESS}^{dis,\max} v_{dis}(t) \quad t \in S_H \quad (8)$$

$$0 \leq P_{BES}^{ch}(t) \leq P_{BES}^{ch,\max} v_{ch}(t) \quad t \in S_H \quad (9)$$

$$v_{dis}(t) + v_{ch}(t) \leq 1 \quad t \in S_H \quad (10)$$

$$S_{OC}(t) = S_{OC}(t-1) - \frac{\eta_{dis} P_{BES}^{dis} \Delta t}{C_{rated}} + \frac{\eta_{ch} P_{BES}^{ch} \Delta t}{C_{rated}} \quad t \in S_H \quad (11)$$

$$S_{OC}^{\min} \leq S_{OC}(t) \leq S_{OC}^{\max} \quad t \in S_H \quad (12)$$

$$\rho_{EV}(t) = K_{EV} \rho_{Grid}(t) \quad t \in S_H \quad (13)$$

In (3), referring to Fig. 2, $\Delta P_{EDR}(t)$ equals to the difference between the EV charging station short-term forecast load $P_{Load}^{Fore}(t)$ and $P_{Load}^{EDR}(t)$. In (4), $\Delta P_{EDR}^{\min}(t)$ is the load reduction signal from the ISO (i.e. minimum decreased load amount participated in the EDR) for the EV charging station. For the EV charging station, V2G is not allowed $P_{Load}^{EDR}(t)$ and $P_{Load}^{EDR}(t)$ is restricted nonnegative in (5). In (6), $P_{EV}(t)$ equals to the sum of $P_{Load}^{EDR}(t)$ and the BES output $P_{BES}(t)$. The BES related constraints are shown in (7)-(12). In (7), $P_{BES}^{dis}(t)/P_{BES}^{ch}(t)$ is the discharge/charge power of the BES. In (8) and (9), $P_{BESS}^{dis,\max}/P_{BES}^{ch,\max}$ is the BES discharge/charge power upper bound, and $v_{dis}(t)/v_{ch}(t)$ is the binary discharge/charge state variable. Constraint (10) will avoid the BES to be operated in charging and discharging modes simultaneously. The state of charge (SOC) of the BES between two adjacent discrete time steps $S_{OC}(t)$ and $S_{OC}(t-1)$ are constrained in (11), where $\eta_{dis}/\eta_{ch}$ is the discharge/charge efficiency parameter. In (12), $S_{OC}^{\min}/S_{OC}^{\max}$ is the lower/upper bound of $S_{OC}(t)$. For different price parameters in the optimization model, to obtain the profit of the EV charging station, $\rho_{EV}(t)$ is usually set higher than $\rho_{Grid}(t)$ ($K_{EV} > 1$ in (13)), while $\rho_{EDR}(t)$ is provided by the ISO.

Therefore, the EDR participation decision making optimization model for the BES-assisted EV charging station is with objective function defined in (1)-(2), and constraints in (3)-(13). As all constraints are linear constraints with binary variables, the formulated EDR participation decision making optimization model is a mixed-integer linear programming (MILP) model, and can be solved by commercial solvers effectively.

*3.2 Decision making process for the BES-assisted EV charging station considering EDR*

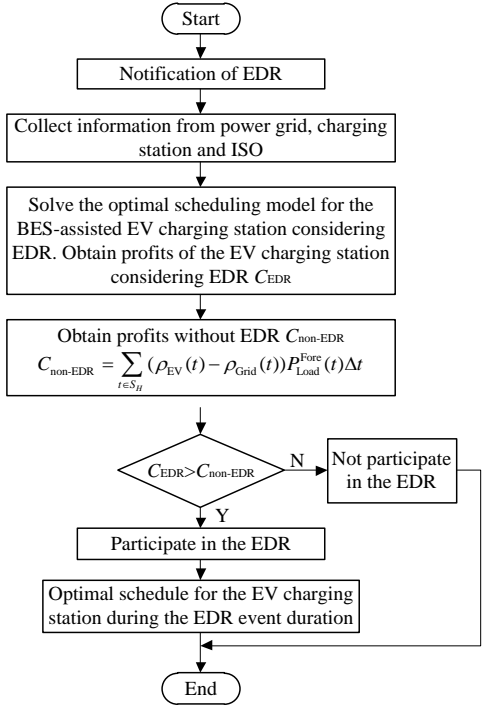

Fig. 4 Flowchart of the decision making process for the BES-assisted EV charging station considering EDR

Based on the optimal scheduling model for the BES-assisted EV charging station considering EDR established in Section 3.1, the detailed flowchart of the decision making process for the BES-assisted EV charging station considering EDR is shown in Fig. 4. Once the notification of EDR from ISO is received, the necessary information from the power grid, charging station and ISO should be collected. Then the profit of the EV charging station considering EDR $C_{\text{EDR}}$ is obtained by solving the optimal scheduling model for the BES-assisted EV charging station considering EDR. The profit of the EV charging station without EDR $C_{\text{non-EDR}}$ is calculated with the formula shown in the Fig.4. The EV charging station decides whether to participate in the EDR or not depends on the values of $C_{\text{EDR}}$ and $C_{\text{non-EDR}}$. If $C_{\text{EDR}} \leq C_{\text{non-EDR}}$, it means that the income from the EDR participation from the ISO cannot compensate the profit loss due to the EV charging load curtailment, and nonparticipation is decided for the EDR.

Otherwise, if $C_{\text{EDR}} > C_{\text{non-EDR}}$, the potential profit loss from the EV charging load curtailment can be over-compensated by the incentive payment for EDR participation from the ISO. It's profitable for the BES-assisted EV charging station to participate in the EDR.

## 4 Case studies

In this section, the validity of the proposed optimal decision making model of BES-assisted EV charging station is demonstrated through case studies. The general input and parameters for the EDR participation decision making optimal model are presented in Section 4.1. The result and analysis of different BES-assisted EV charging station EDR decision making cases are presented in Section 4.2. Comparisons of different BES-assisted EV charging station EDR decision making cases are included in Section 4.3.

*4.1 General inputs and parameters for the EDR participation optimal decision making model*

A test BES-assisted EV charging station is utilized in the case studies. The EV charging station short-term forecast load and the TOU electricity price for a whole typical day [25] are shown in Fig. 5 and utilized as inputs for the EDR participation optimal decision making model.

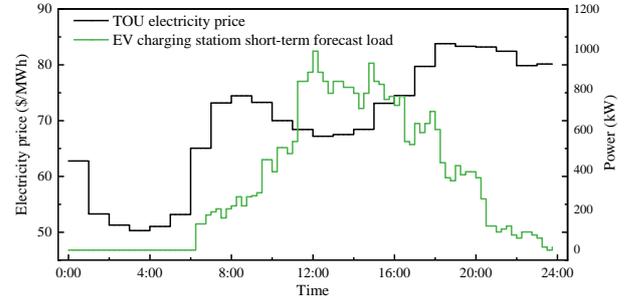

Fig. 5 EV charging station short-term forecast load and the TOU electricity price for a whole typical day

Table 1 Parameters for the EDR participation optimal decision making model

| Parameter | Value | Parameter | Value |
| --- | --- | --- | --- |
| $H$ | 5 hour | $\Delta P_{\text{EDR}}^{\min}(t)$ | 150 kW |
| $\Delta t$ | 15 min | $P_{\text{BESS}}^{\text{dis,max}} / P_{\text{BES}}^{\text{ch,max}}$ | 55/50 kW |
| $K_{\text{EV}}$ | 3 | $\eta_{\text{dis}} / \eta_{\text{ch}}$ | 1.15/0.85 |
| $C_{\text{rated}}$ | 400 kWh | $S_{\text{OC}}^{\min} / S_{\text{OC}}^{\max}$ | 0.2/0.85 |

The main parameters for the EDR participation optimal decision making model are tabulated in Table 1. Except for the paramaters included in Table 1, other essential paratemters e.g. the specific EDR event duration, the EDR participation incentive payment price $\rho_{\text{EDR}}(t)$, and the initial

SOC of the BES at the beginning of the EDR event duration $S_{OC}(t_s-1)$ etc. for different BES-assisted EV charging station EDR decision making cases will be presented in Section 4.2.

*4.2 Result and analysis of different BES-assisted EV charging station EDR decision making cases*

A. Case 1

In Case 1, the EDR notification moment is set at 6:00, and after 1 hour EDR decision making duration, the study EDR event duration in Case 1 is 7:00-12:00. Other parameters for the EDR participation optimal decision making model of Case 1 are listed in Table 2.

Table 2 Other parameters for the EDR participation optimal decision making model of Case 1

| Parameter | Value |
|---|---|
| EDR notification moment | 6:00 |
| EDR decision making duration | 1 hour |
| EDR event duration | 7:00-12:00 |
| $\rho_{EDR}(t) \ t \in S_H$ | 75 $/MWh |
| $S_{OC}(t_s-1)$ | 0.85 |

Table 3 Solving results of the decision making process for the BES-assisted EV charging station considering EDR of Case 1.

| Index | Without EDR | With EDR |
|---|---|---|
| Station profit ($C_{non-EDR}/C_{EDR}$) | $303.69 | $300.37 |
| Income from the EV charging loads EV load | $455.54 | $342.07 |
| Income from the EDR participation | $0 | $56.25 |
| Cost for the electricity purchased from the grid | $151.84 | $97.95 |
| Decision of EDR participation | Nonparticipation | |

Solving results of the decision making process for the BES-assisted EV charging station considering EDR of Case 1 are shown in Table 3. The key index values in the third column in Table 3 correspond to the solving results of the optimal scheduling model for the BES-assisted EV charging station considering EDR. As shown in Fig. 5, the EV charging station short-term forecast load keeps increasing during the EDR event duration 7:00-12:00, and the potential profit loss because of EV charging load curtailment is very high. Meanwhile, referring to Table 2, the EDR participation incentive payment price is relative low (75 $/MWh). So the income from the EDR participation from the ISO cannot compensate the profit loss due to the EV charging load curtailment in Case 1. The total profit of the EV charging station considering EDR $C_{EDR}$ ($300.37 in Table 2) is less than the profit without EDR $C_{non-EDR}$ ($303.69 in Table 2).

According to the decision making process in Section 3.2, the BES-assisted EV charging station decides nonparticipation in the EDR notification at 6:00 in Case 1.

B. Case 2

In Case 1, the EDR notification moment is set at 15:00, and after 1 hour EDR decision making duration, the study EDR event duration in Case 2 is 16:00-21:00. Other parameters for the EDR participation optimal decision making model of Case 2 are listed in Table 4.

Table 4 Other parameters for the EDR participation optimal decision making model of Case 2

| Parameter | Value |
|---|---|
| EDR notification moment | 15:00 |
| EDR decision making duration | 1 hour |
| EDR event duration | 16:00-21:00 |
| $\rho_{EDR}(t) \ t \in S_H$ | 200 $/MWh |
| $S_{OC}(t_s-1)$ | 0.85 |

Table 5 Solving results of the decision making process for the BES-assisted EV charging station considering EDR of Case 2.

| Index | Without EDR | With EDR |
|---|---|---|
| Station profit ($C_{non-EDR}/C_{EDR}$) | $490.42 | $625.38 |
| Income from the EV charging loads EV load | $735.63 | $297.99 |
| Income from the EDR participation | $0 | $408.50 |
| Cost for the electricity purchased from the grid | $245.21 | $81.11 |
| Decision of EDR participation | Participation | |

Solving results of the decision making process for the BES-assisted EV charging station considering EDR of Case 2 are shown in Table 5. In Fig. 5, the EV charging station short-term forecast load gradually decreases during the EDR event duration 16:00-21:00, and due to the relative high EDR participation incentive payment price (200 $/MWh), the potential profit loss because of EV charging load curtailment can be over-compensated by the incentive payment for EDR participation from the ISO. As $C_{EDR}$ ($625.38 in Table 5) is larger than the $C_{non-EDR}$ ($490.42 in Table 5), the BES-assisted EV charging station decides participation in the EDR notification at 15:00 in Case 2.

*4.3 Comparisons of different BES-assisted EV charging station EDR decision making cases*

The brief comparisons of the total profit of the EV charging station with/without EDR are shown in Fig. 6. In different BES-assisted EV charging station EDR decision making cases with different inputs and parameters, it's not always profitable for the EV charging station to participate in the EDR. The comparisons of different BES-assisted EV charging station EDR decision making cases in Section 4.2 can demonstrate that through the optimal decision making model proposed in Section 3, correct and profitable EDR participation decision can be determined for the BES-assisted EV charging station effectively.

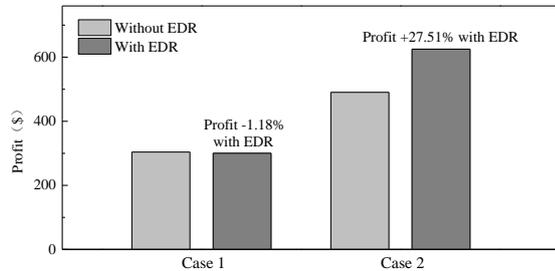

Fig. 6 Brief comparisons of the total profit of the EV charging station with/without EDR

Table 6 Total profit of the EV charging station considering EDR in Case 2 with different BES capacity parameters

| BES capacity | Percentage of the BES capacity in Case 2 | Total profit of the EV charging station considering EDR |
| --- | --- | --- |
| 0 kWh | 0% $C_{rated}$ | $570.71 |
| 80 kWh | 20% $C_{rated}$ | $581.82 |
| 160 kWh | 40% $C_{rated}$ | $592.89 |
| 240 kWh | 60% $C_{rated}$ | $603.92 |
| 320 kWh | 80% $C_{rated}$ | $614.80 |
| 400 kWh | 100% $C_{rated}$ | $625.38 |
| 480 kWh | 120% $C_{rated}$ | $635.26 |
| 560 kWh | 140% $C_{rated}$ | $637.45 |
| 600 kWh | 150% $C_{rated}$ | $637.45 |
| 800 kWh | 200% $C_{rated}$ | $637.45 |

Further analyse the effect of different BES capacity parameters to the total profit of the EV charging station considering EDR through changing the BES capacity in Case 2 in Section 4.2. Table 6 shows the total profit of the EV charging station considering EDR in Case 2 with different BES capacity parameters. Referring to Table 6, in general larger BES capacity contributes to higher total profit of the EV charging station considering EDR. But in Table 6, when the BES capacity reaches 140% $C_{rated}$ (560 kWh), the BES capacity is saturated to further improve the total profit of the EV charging station considering EDR. Comparison results in Table 6 demonstrate that the local BES in the EV charging station can improve the charging station's ability to participate in the EDR. And under certain limits, enlarging the BES capacity can lead to higher total profit of the EV charging station considering EDR.

## 5 Conclusion

In this paper, an optimal decision making model of BES-assisted EV charging station considering incentive demand response is proposed. The EDR is considered as the representative incentive demand response in this paper. The methodology for BES-assisted EV charging station participated in EDR is presented. The optimal scheduling model for the BES-assisted EV charging station considering EDR is established. And on this basis, the detailed decision making process for the BES-assisted EV charging station considering EDR is proposed.

The validity of the proposed optimal decision making model of BES-assisted EV charging station is demonstrated through case studies. The comparison results in the case studies demonstrate that : 1) Through the optimal decision making model, correct and profitable EDR participation decision can be determined for the BES-assisted EV charging station effectively. 2) Local BES in the EV charging station can improve the charging station's ability to participate in the EDR.